\documentclass[prl,aps,twocolumn,showpacs]{revtex4}
\usepackage{epsfig}
 
\begin{document}

\title{Creation of macroscopic quantum superposition
states by a measurement}
\author{I.E. Mazets$^{1,2}$, G. Kurizki$^3$, M.K.  Oberthaler$^4$
and J. Schmiedmayer$^2$}
\affiliation{$^1$ A.F. Ioffe Physico-Technical Institute,
194021 St.Petersburg, Russia, \\
$^2$ Atominstitut der \"Osterreichischen Universit\"aten, 1020
Vienna, Austria, \\
$^3$ Chemical Physics Dept., Weizmann Institute of Science, 76100
Rehovot, Israel, \\
$^4$ Kirchhoff Institut f\"ur Physik, Universit\"at
Heidelberg, 69120 Heidelberg, Germany}

\begin{abstract}
 We propose a novel protocol for the creation of
macroscopic quantum superposition (MQS) states
based on  a measurement of a non-monotonous function of a quantum collective
variable. The main advantage of this protocol is that it does not require
switching on and off nonlinear interactions in the system.
We predict this protocol to allow the creation of multiatom MQS by
measuring the number of atoms coherently outcoupled from a two-component
(spinor) Bose-Einstein condensate.
\end{abstract}


\maketitle

Macroscopic quantum superposition (MQS) states
embody the famous paradox formulated
by Schr\"odinger \cite{sch35} that quantum mechanics
admits the existence of a cat in a quantum superposition of ``dead'' and
``alive'' states. MQS states are not only interesting
from the fundamental point of view, but are  also promising for many
applications, such as precision quantum measurements \cite{bl} or
quantum computing \cite{lb}. Most of the methods proposed or used so far
\cite{har,wine,zio,sq,ys} for MQS state realization have been based on
nonlinear unitary dynamics.

MQS states have not yet been  realized in an
atomic Bose-Einstein condensate (BEC), although its coherently coupled
components (in a double trap or in a spinor condensate) can be
isomorphous to a Josephson junction \cite{go} with very promising
properties for the realization of MQS states. Yet such realization may be
impeded by difficulties inherent in the proposed methods, based on the
nonlinear dynamics of the BEC, either isolated from the environment
\cite{dnm,mpr,zl}, or perturbed by it in a specific (``symmetrized'')
way \cite{zurek}.

We propose an alternative generic method of MQS state creation,
wherein the macroscopic quantum system is linear, but the
non-linearity is introduced by the measurement process
\cite{Rfn14}. Namely,
assume that the system is initially in the quantum state
$\Psi _{in}(x)$
characterized by a wide spread of the macroscopic collective variable
$x$. If we measure not $x$ itself, but a certain non-monotonous
function $f(x)$,  then the
state of the system collapses to
\begin{equation}
\Psi _{out}(x)=\Psi _{in}(x)g[f(x)-f_0],
\label{eq:1}
\end{equation}
where $f_0$ is the measurement outcome and
$g(y)$ is a narrow-peaked function centered at $y=0$. In particular,
if the Hamiltonian of the system-detector interaction is
$\hbar \hat{P}f(x)$,
$\hat{P}$ being the detector's momentum operator, then
$g[f(x)-f_0]=
\langle f_0| \exp [-i\hat{P}f(x)T_m] |\textrm{i}\rangle  $,
where $|\textrm{i}\rangle $ and $|f_0\rangle $ are the detector states
before and after the measurement, respectively, and $T_m$ is the
duration of the measurement. If the
equation $f(x)=f_0$ has multiple roots, well separated from each other,
then Eq.~(\ref{eq:1}) describes a macroscopic superposition state. For the
sake of definiteness, assume that $f(x)$ is even, and $x=\pm x_0$ are
the roots of the equation $f(x)=f_0$. Then, apart from
a normalization factor, Eq.~(\ref{eq:1}) reduces to
\begin{eqnarray}
\Psi _{out}(x)&\propto &\Psi _{in} (x_0)g[f'(x_0)(x-x_0)]+
\nonumber \\ && \Psi (-x_0)g[f'(-x_0)(x+x_0)].
\label{eq:2}
\end{eqnarray}
The two peaks are well-resolved if $g[x_0f'(x_0)]\ll g(0)$.
The generalization of the derivation above to the density matrix
formalism is straightforward but cumbersome.

The cardinal question is how to realize the necessary measurement.
As a particular, experimentally relevant example, it will be shown
that such a measurement is realizable on outcoupled atoms
in the regime of coherent population
trapping \cite{AM} in a two-component BEC.
To this end, consider a trapped BEC of atoms in
two sublevels, $|1\rangle $ and $|2\rangle $, of the atomic ground state.
The conjugate collective variables are the
intercomponent atom-number difference  and the relative phase
$\phi $, which are analogous to the Josephson junction conjugate
variables, just like their counterparts in a double-well BEC
\cite{s97,go}.   We coherently couple
the levels 1 and 2 to a common third level 0,
thus forming a $\Lambda$-scheme of excitation. If the intercomponent phase
is well-defined then the population of the state $|0\rangle $ is
proportional to $V_1^2N_1+V_2^2N_2+2V_1V_2\sqrt{N_1N_2}\cos \phi $,
$V_\beta $ being
the (real) Rabi frequency for the $\beta  -0$ transition,
[see inset to Fig. 1(a)] and $N_\beta $ being the mean atom number in the
level  $\beta  $, $\beta =1,\, 2$. If, instead,  $\phi $ is uncertain,
then we can expect that measuring the population
of the level 0 (or its growth rate) will yield a certain value of
$\cos \phi =\cos \phi _0$ and thereby
project the state of the system to the MQS Eq. (\ref{eq:2})
where $\phi $ stands for $x$.

The Hamiltonian in the interaction representation reads as
\begin{equation}
\hat{H}=i\hbar [ V_1 (\hat{b}_0^\dag \hat{b}_1 -
\hat{b}_1 ^\dag \hat{b}_0 ) + V_2 (\hat{b}_0^\dag \hat{b}_2 -
\hat{b}_2 ^\dag \hat{b}_0 )] .
\label{eq:4}
\end{equation}
It is convenient to parametrize the Rabi frequencies as
$V_1=V\cos \alpha $, $ V_2=V\sin \alpha $,
$V =(V_1^2+V_2^2)^{1/2}  $.
 
Consider first the coherent dynamics generated by Eq. (\ref{eq:4}),
assuming for simplicity that
initially (at $t=0$) the BEC is in the product of Fock states
of the components $\beta =1, \, 2$:
\begin{equation}
|\Psi _{in}^{N_1\, N_2} \rangle
=(N_1!N_2!)^{-1/2}\hat{b}_1^{\dag \, N_1}
\hat{b}_2^{\dag \, N_2} |\mathrm{vac} \rangle ,
\label{eq:3}
\end{equation}
where $|\mathrm{vac}\rangle $ is the vacuum of the atomic field and
$\hat{b}_\beta ^{\dag }$ is the creation operator for an atom in the
respective internal state $|\beta \rangle $
and in the lowest-energy motional state of the trap. At time $t$ the
system evolves into the state
\begin{eqnarray}
|\Psi (t)\rangle &=&(N_1!N_2!)^{-1/2} (\xi _1\hat{b}_1^\dag +
\xi _2\hat{b}_2^\dag +\xi _0 \hat{b}_0^\dag )^{N_1}\times
\nonumber \\ &&
(\xi _2\hat{b}_1^\dag + \tilde{\xi }_1\hat{b}_2^\dag +
\tilde{\xi }_0 \hat{b}_0^\dag )^{N_2}|\mathrm{vac}\rangle ,
\label{eq:6}
\end{eqnarray}
where $\xi _1=\cos ^2\alpha \cos Vt+\sin ^2\alpha $, $\xi _2 =
\cos \alpha \sin \alpha (\cos Vt -1)$, $\xi _0=\cos \alpha \sin Vt$,
and $\tilde {\xi }_\beta $ is obtained from $\xi _\beta $ by changing
$\alpha $ to $\frac \pi 2 -\alpha $.
We will see later that accurate knowledge of 
$N_1$ and $N_2$ (or, at least, their difference) 
is essential for the {\em detection} of the resulting MQS state.

Here and in what follows we assume that the number $n_0$ of 
atoms outcoupled to the level 0 is always much less than 
$ N_1, N_2$, therefore the
depletion of $N_1$ and $N_2$ in the course of the evolution can be
neglected. Also, to make the expressions less
cumbersome, we assume $N_2\approx N_1\equiv N$. The corresponding
eigenstates, defined for the subsystem of atoms in the levels 1 and 2
only, are denoted by $|j\rangle _{12}$.
Upon assuming equal Rabi frequencies, $V_1=V_2$,
Eq.~(\ref{eq:6}) reduces to
\begin{eqnarray}
|\Psi (t)\rangle &=&\frac 1{2^N} \sum _{j=0}^N \sum _{n_0=0}^{2j}
\frac {(-1)^j(2j)!\sqrt{(2N-2j)!}}{j!(N-j)!\sqrt{n_0!(2j-n_0)!}}
\times \nonumber \\ &&
\cos ^{2j -n_0}Vt \sin ^{n_0}Vt|j\rangle _{12}|n_0\rangle _0 ,
\label{eq:9} \\
|n_0\rangle _0&=&(n_0!)^{-1/2}\hat{b}_0^{\dag \, n_0}
|\mathrm{vac}\rangle .
\label{eq:9-bis}
\end{eqnarray}

The level 0 populated by $n_0$ atoms has a finite coherence time,
caused by spontaneous relaxation, if it is
optically excited, or by the translational motion of atoms, if
it is magnetically untrapped. Most importantly,
its coherence is limited by the rate of the measurements
that reveal the quantum information needed to project the initial
state (\ref{eq:3}) onto a MQS. Depending on the ratio of the
characteristic time $V^{-1}$ of the evolution under Hamiltonian
(\ref{eq:4}) to the coherence time, the dynamics may be anywhere
between two limiting regimes.
If the Hamiltonian evolution is much faster than the coherence
time, we approach the limit of a
fully coherent regime. The opposite case corresponds
to the regime of continuous observation. Our results
in the limit of continuous observation bear
similarity  to those of Ref. \cite{cd97}, which is a stationary
analysis of  the relative phase of two independent condensates
interfereing via a beam-splitter under idealized conditions.
Related measurement-based schemes to create MQS states have been 
suggested in Refs. \cite{Ref19a,NB}. 

  \begin{figure}

  \begin{center}
  \centerline{\epsfig{file=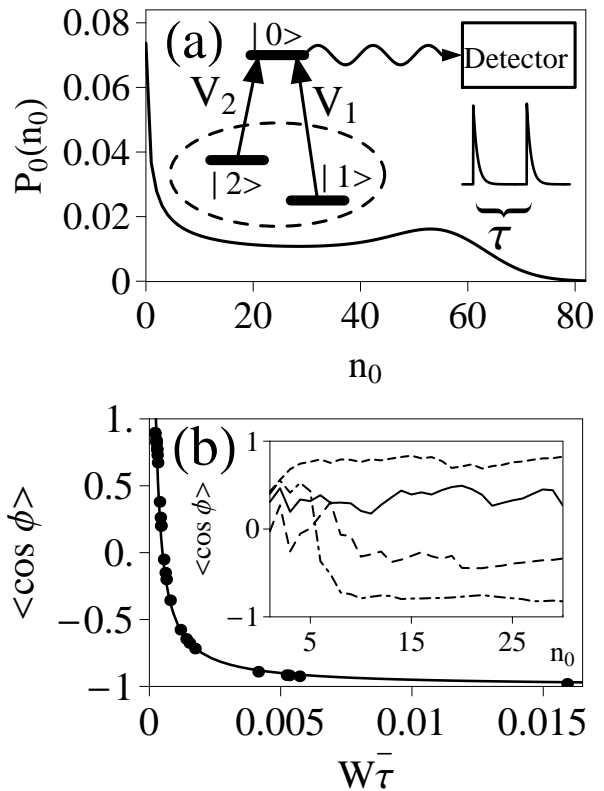,width=7.8cm}}
  \end{center}

  \caption{
  (a) Probability distribution of the $|0\rangle $ state population in
  the coherent regime. $N_1=N_2=1000$, $\alpha =\pi /4$,
  and $\langle n_0\rangle =30$. The units on this plot and
  the subsequent plots are dimensionless. Inset: schematic
  representation of a measurement. The sequence of detected atoms can
  correspond to either the coherent regime (overlapping signals, i.e.,
  simultaneous detection of $n_0$ atoms) or to continuous observation
  (time-resolved signals). (b) Regime of continuous
  observation for the same $N_1$, $N_2$, and $\alpha $.
  Dots: $\langle \cos \phi \rangle $ after outcoupling of 30 atoms as
  a function of the mean time interval $\bar{\tau }$ between detection
  events for different QMC runs.
  Solid line: approximation by $\bar{\tau }
  =[2WN(1+\langle \cos \phi \rangle )]^{-1}$.
  Inset: Four typical histories of establishing a definite value of
  $\langle \cos \phi \rangle $, as the number $n_0$ of outcoupled atoms
  grows. }
  \label{fig_tcfg1}
  \end{figure}

In the limit of a fully coherent regime, a single, instantaneous
measurement yields the number $n_0$ of atoms in the level 0.
If we drop the summation over $n_0$ in Eq. (\ref{eq:9}), we obtain
(in the unnormalized form) the state to which the system collapses
upon detecting exactly $n_0$ atoms in the state $|0\rangle $.
Such a measurement projects the state (\ref{eq:9}) onto an
eigenstate of the phase-cosine operator
$\cos \hat{\phi }=( 2\sqrt{N_1N_2}) ^{-1}
(\hat{b}_1^\dag \hat{b}_2+\hat{b}_2^\dag \hat{b}_1 ) $,
whose spectrum is discrete \cite{dissp}:
$\cos \phi =-1+ {2j}/N$,  $j =0,\, \frac 12,\, 1,\,
\frac 32,\, \dots \, ,\, N$.
Note that terms with half-integer $j$'s are absent in Eq. (\ref{eq:9}).
The crux of our method is that even if $\phi $ is not directly measurable,
$\cos \phi $ is.  The corresponding conditional probability distribution
of the relative
phase, can be fairly approximated (if $\phi $ is not too close to 0 or
$\pm \pi $) by
\begin{equation}
P_{ph}(\phi |n_0)=C_{ph}\exp \left \{
-\frac {[n_0-N\sin ^2Vt(1+\cos \phi )]^2}
{2n_0} \right \} ,
\label{eq:10}
\end{equation}
$C_{ph}$ being the normalization factor. As follows from
Eq. (\ref{eq:9}), the state corresponding to the double-peaked probability
distribution (\ref{eq:10}) is a pure state, which is therefore a MQS.
However, such a MQS is not a sum of two Gaussian wave packets
(harmonic-oscillator coherent states), as might be naively expected,
but has far more complicated form, since {\em terms of
alternating sign} appear in the r.h.s. of Eq. (\ref{eq:9}). Such a
behaviour precludes taking the limit $N\rightarrow \infty $ in
Eq. (\ref{eq:9}). However, this can be done for
$P_{ph}(\phi |n_0)$, if we simultaneously set $t\rightarrow 0$,
keeping $\langle n_0\rangle =\mathrm{const}$.

Upon tracing out atoms in the levels 1 and 2, we
obtain the probability distribution $P_0(n_0)$ of the
population of the level 0. This distribution strongly differs from
a Poissonian form and is
plotted in Fig.~1(a). For $n_0\, ^>_\sim \, 3$ 
it is excellently fit by its
quasicontinuous limit, whose analytic expression is cumbersome and
will be given elsewhere. The mean value and dispersion of $n_0$ are,
respectively, $\langle n_0\rangle =N\sin ^2Vt $ and $\langle n_0^2
\rangle -\langle n_0\rangle ^2 =N\sin ^2Vt\cos ^2Vt +\frac 12 N^2
\sin ^4Vt$.

The regime of continuous observation takes place if $V\ll \gamma $,
where $\gamma $ is the inverse coherence time of the level 0.
In this case the atoms in the level 0 are detected
one by one. The information on
the value of $\cos \phi $ is revealed by measuring the time
intervals $\tau $ between consecutive atom detection events.

We have constructed a quantum Monte Carlo (QMC) algorithm that simulates
individual measurement outcomes. The QMC \cite{AL} method allows one to
simulate the emergence of an interference pattern with {\em a priori}
unknown phase for two spatially interfering, dissipative BECs, each being
initially in a Fock (number) state \cite{jy}.

  \begin{figure}

   \begin{center}
  \centerline{\epsfig{file=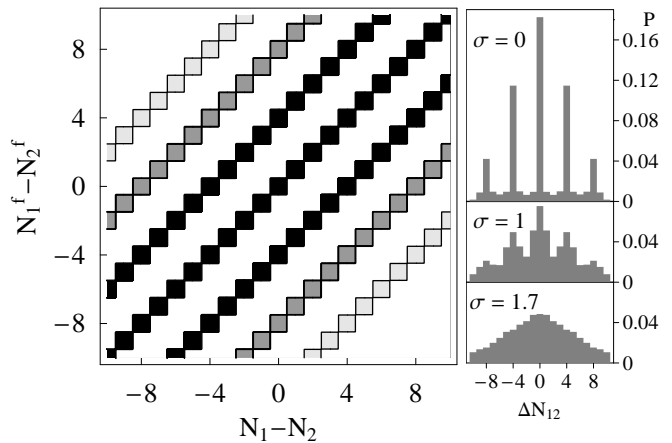,width=\columnwidth}}
  \end{center}

  \caption{
  Left panel: Distribution of the final difference $N_1^f-N_2^f$
  as a function of the initial number difference $N_1-N_2$
  (the darkest area of this contour plot corresponds to the
  highest probability) for $N_1\approx N_2\approx 10^3$. Right panel:
  Probability distribution $P$ of  $\Delta N_{12}$ 
  (final-state number difference
  centered at the initial state number difference),
  assuming  Poissonian distributions of the
  initial atom numbers around mean values
  $\bar{N}_1=1000$ and
  $\bar{N_{2}}=1000$.   The plots are labeled by the total r.m.s.
  error $\sigma $ of the number measurement. For both panels the
  number of outcoupled atoms $\nu =26$ and $\langle \cos \phi
  \rangle =0$.   }
  \label{fig_dgr2}
  \end{figure}

In Fig. 1(b) we present our QMC simulation of the
outcoupling of $\nu $ atoms to the level 0, starting from the
state (\ref{eq:3})  and detecting them one by one. The
output of the numerical simulations was twofold: Firstly, we
obtained the set of time intervals $\tau _i$, $i=1,\, 2,\, \dots ,
\, \nu $, between consecutive detection events. Secondly, we
monitored the state of the remaining atoms after each 
outcoupling event. The average time interval
$\bar{\tau }$ was found to be in  very good agreement
with its estimation $[2WN(1 +
\langle \cos \phi \rangle )]^{-1}$, where $\langle \cos \phi
\rangle $ is the mean value of the $\cos \hat{\phi }$
operator in the state emerging after the outcoupling of the last
atom in the measured sequence
and $W=V^2/\gamma $. The statistical distribution
of sequences of $\tau _i$ in
each QMC run was found  to be very close to
Poissonian, with the probability density $\bar{\tau }^{-1}
\exp (-\tau /\bar{\tau } )$. This implies that the first few
measurements of $\tau $ determine quite  well the value of
$\langle \cos \phi \rangle $ and, hence, the subsequent evolution
of the system [see inset to Fig.~1(b)].

To prove the MQS nature of the final state, one has to observe the
interference structure in the distribution of the final
atom-number difference, the variable conjugate to $\phi $. To this end, 
we plot in Fig.~2 the probability distribution $P(\Delta N^f)$ of
atom number difference in the final state $\Delta N^f = N_1 ^f-N_2^f$, 
as a function of the initial number difference $\Delta N =
N_1-N_2$.  The peaks in the distribution of $\Delta N^f$ are
separated by {\em four} atom counts, thus indicating 
interference effects in the MQS state. The position of these
interference peaks depends on the initial number difference, as
shown in Fig.~2 (left panel).

If we start with a mixed initial state, characterized by
independent Poissonian fluctuations of $N_1$, $N_2$ around their
mean values $\bar{N}_1$, $\bar{N}_2$, then the interference
patterns can be recognized by looking at the quantity
$\Delta N_{12}=N_1^f -N_2^f-(N_1-N_2)$, the final atom-number
difference centered at the initial number difference.  This
interference structure  still persists for a measurement error of 1
atom in determining $\Delta N_{12}$,  $\sigma(\Delta N_{12}) =1.0$,
but disappears for $\sigma(\Delta N_{12}) =1.7$.

Such a high sensitivity of the MQS state detection  to the
accuracy of counting the atoms seems to be a common
feature of measurement-based 
MQS phase-state creation methods (cf. Ref.\cite{NB}).
Non-demolition measurements of atom-number differences in high-$Q$ 
cavities \cite{QND} or quantum culling techniques \cite{cull}
may help satisfy this stringent requirement.

There are different possibilities to realize
the $\Lambda $-scheme excitation. Preferably, the level 0 is to be
excited via coupling to the ground-state sublevels by laser-induced
single-photon transitions. In this case, to avoid complications resulting
from co-operative, bosonic-enhanced relaxation \cite{jcr}, we assume that
$N_1+N_2\ll (k_LR)^2$, where $k_L$ is the laser radiation wavenumber and
$R$ is the BEC size.
The BEC should be collisionally thin, i.e., the number of collisions
per atom moving with the momentum
$\sim \hbar k_L$ imparted by a scattered photon should be negligible.

Our method can be  efficiently implemented with $^{87}$Rb
condensates. The triplet and singlet $s$-wave scattering
lengths are very close to each other for this isotope, and therefore
inter- and intracomponent scattering length difference for $^{87}$Rb
is of about 1~\% \cite{Rba}. This provides several advantages. Firstly, 
quantum diffusion of $\phi $ due to slight difference in the
mean-field interaction energies for the different states will
destroy the intercomponent phase coherence on a
very long time scale $\sim 1$~s,
if we assume the following parameters: total number of atoms is
about 2000, the BEC size $R\approx 10~\mu $m.
Secondly, the inelastic losses
in $^{87}$Rb are suppressed due to the closeness of the triplet and
singlet scattering lengths \cite{Rba} and can be totally neglected
for the given BEC parameters on time scales up to 0.1~s. Additionally,
in such a small BEC, bosonic enhancement \cite{jcr} does not
significantly affect the relaxation of the optically excited state, and
the probability of a collision for an atom that acquires recoil velocity
after scattering a photon is of the order of 0.01. Hence, a few
dozens of atoms can be outcoupled without destroying the remaining
BEC by collisions.

The collisions with thermal atoms and scattering of resonance stray 
photons \cite{dnm} destroy the macroscopic coherence. In general, the 
rate for the MQS destruction is $(N_1+N_2)$ faster than the rate of 
excitation of an individual atom from the condensate (we assume that 
every such scattering is energetic enough to change the translational 
state of atomic motion). Since the lifetimes of a BEC of the order of 
several tens of seconds are experimentally feasible, an MQS can persist 
in a system with $N_1\approx N_2\sim 10^3$ for tens of milliseconds, 
which is long enough to perform the necessary optical manipulation and 
detection steps whose duration lies in the sub-millisecond range. 

To conclude, we have proposed a novel method for creating 
macroscopic superposition states in two-component BECs by measuring
the cosine of the intercomponent phase. Either 
``snapshot'' detection of the number of
outcoupled  atoms or their continuous observation 
have been shown to yield
the desired result (Fig.~1). This method allows for 
fluctuations of the atom-number difference in
the initial state, if the error in atom counting by 
a detector is maintained at the level of $\pm 1$ atom (Fig.~2). Each
measurement of $\cos \phi $ yields a MQS, provided that the measured
$\phi $ is not too close to 0 or $\pm \pi $. In particular, for the 
parameters of Fig. 2 
(right panel), the peaks in two-peaked phase distribution 
appear to be well-resolved if $| \langle \cos \phi \rangle | < 0.3$. 
This distinguishes
our method from the method of optical MQS
generation recently developed \cite{ourj} for homodyne measurements of
quantum field quadratures that are suitable for photons, but not for
atoms. A major advantage of our method compared to methods based on
nonlinear interactions in the system itself \cite{dnm,mpr,zl,zurek} is
that it allows one to create MQS rapidly enough, regardless of the
nonlinearity smallness, well before the interaction with the
environment brings about the collapse of the macroscopic superposition.

This work is supported by the German-Israeli Foundation,
the EC (MIDAS STREP) and INTAS
(project 06--1000013--9427). I.E.M. acknowledges the Lise Meitner
Fellowship by the FWF.


\begin{thebibliography}{99}
\bibitem{sch35} {\sc{Schr\"odinger  E.}}, {\em Naturwiss}. {\bf 23} (1935) 807;
{\em ibid.,} 823; {\em ibid.,} 844.

\bibitem{bl} {\sc{ Bollinger J.J.,  Itano W.M.,  Wineland D.J., and 
Heinzen D.J.,}} {\em Phys. Rev. A} {\bf 54} (1996) R4649.

\bibitem{lb} {\sc Liebfried D.} {\em et al.}, {\em Nature} {\bf 438} (2005) 639. 

\bibitem{har} {\sc  Brune M.} {\em et al.}, {\em Phys. Rev. Lett.}
{\bf 77} (1996) 4887.

\bibitem{wine} {\sc Monroe C.,  Meekhof D.M.,  King B.E., and  Wineland D.J.},
{\em Science} {\bf 272}  (1996) 1131.

\bibitem{zio} {\sc  Friedman J.R.,  Sarachik M.P.,  Tejada J., and Ziolo R.},
{\em Phys. Rev. Lett}. {\bf 76} (1996) 3830.

\bibitem{sq} {\sc Rouse R.,  Han S., and  Lukens J.E.}, {\em Phys. Rev. Lett}.
{\bf 75} (1995) 1614; {\sc Nakamura Y.,  Pashkin Y.A., and  Tsai J.S.},
{\em Nature } {\bf 398} (1999) 786;
{\sc Friedman J.R.} {\em et al.}, {\em Nature} {\bf 406} (2000) 43.

\bibitem{ys} {\sc Yurke B. and  Stoler D.}, {\em Phys. Rev. Lett.}
 {\bf 57}  (1986) 13.

\bibitem{go} {\sc Albiez M.} {\em et al.}, {\em Phys. Rev. Lett.} {\bf 95}, 
(2005) 010402; {\sc   Gati R. and  Oberthaler M.K.}, {\em J. Phys. B} {\bf 40}
(2007) R61.

\bibitem{dnm} {\sc  Huang Y.P. and  Moore M.G.}, {\em Phys. Rev. A} {\bf 73}
(2006) 023606.

\bibitem{mpr}  {\sc Mahmud K.W.,  Perry H., and  Reinhardt W.P.}, {\em J. Phys. B}
{\bf 36} (2003) L265.

\bibitem{zl} {\sc Micheli A., Jaksch D., Cirac J.I., and  Zoller P.},
{\em Phys. Rev. A} {\bf 67} (2003) 013607.

\bibitem{zurek} {\sc  Dalvit D.A.R.,  Dziarmaga J., and  Zurek W.H.},
{\em Phys. Rev. A } {\bf 62} (2000) 013607.

\bibitem{Rfn14} {\sc Massar S. and Polzik E.S.}, {\em Phys. Rev. Lett.} 
{\bf 91} (2003) 060401. 

\bibitem{AM} {\sc E. Arimondo}, in {\em Progress in Optics}, ed. E. Wolf,
vol. 35, Elsevier, Amsterdam (1995), p. 257.

\bibitem{s97} {\sc  Smerzi A.,  Fantoni S., Giovanazzi S., and Shenoy S.R.},
{\em Phys. Rev. Lett.} {\bf 79} (1997) 4950.


\bibitem{cd97} {\sc  Castin Y. and Dalibard J.}, {\em Phys. Rev. A} {\bf 55}
(1997) 4330.

\bibitem{Ref19a} {\sc  Ruostekoski J.,  Collett M.J.,  Graham R., and 
Walls D.F.}, {\em Phys. Rev. A } {\bf 57} (1998) 511.  

\bibitem{NB} {\sc Dunningham J.A.,  Burnett K.,  Roth R.,
and  Phillips W.D.}, {\em New J. Phys.}, {\bf 8} (2006) 182.

\bibitem{dissp} {\sc Pegg D.T. and  Barnett S.M.}, {\em Europhys. Lett}.
{\bf 6} (1988) 483; {\sc  Luis A. and  S\'anchez-Soto L.L.},
{\em Phys. Rev. A } {\bf 48} (1993)  4702.
 

\bibitem{AL} {\sc Morigi G.,  Zambon B.,  Leinfellner N. and
 Arimondo E.}, {\em Phys. Rev. A } {\bf 53} (1996) 2616.

\bibitem{jy} {\sc  Javanainen J. and Yoo S.M.}, {\em Phys. Rev. Lett.}
{\bf 76} (1996) 161.
 

\bibitem{QND}
If one tunes a high-$Q$ cavity resonance inbetween the two hyperfine
ground states, then one can directly measure the atom-number difference
between the two states in the 
strong coupling regime of cavity-trapped atoms, see: {\sc F. Brennecke}
{\em et al.}, {\em Nature} {\bf 450} (2007) 268; {\sc  Colombe Y.} {\em et
al.}, {\em Nature } {\bf 450} (2007) 272. 


\bibitem{cull} {\sc Dudarev A.M.,  Raizen M.G., and Qian Niu},
{\em Phys. Rev. Lett.} {\bf 98} (2007) 063001.

\bibitem{jcr} {\sc Javanainen J.}, {\em Phys. Rev. Lett}. {\bf 72} (1994) 2375.

\bibitem{Rba} {\sc Burke J.P., Jr.,  Bohn J.L.,  Esry B.D., and 
Greene C.H.}, {\em Phys. Rev. Lett.} {\bf  80} (1998) 2097.

\bibitem{ourj} {\sc Ourjoumtsev A.,  Jeong H.,  Tualle-Brouri R. and
Grangier P.}, {\em Nature} {\bf 448} (2007) 784.


\end{thebibliography}
\end{document}